\documentstyle[12pt,eqnus]{article}

\title{Anyonic Construction of the $sl_{q,s}(2)$ Algebra}
\author{$\mbox{J.L. Matheus--Valle }^{a,b}$
  and  $\mbox{M. R-Monteiro }^{a}$ \\    \\
        $^a$ CBPF/CNPq, Rua Dr. Xavier Sigaud, 150 \\
         22290-180 Rio de Janeiro, RJ, Brazil \\  \\
        $^b$ Departamento de F\'\i sica, ICE \\
         Universidade Federal de Juiz de Fora \\
          36000-000 Juiz de Fora, MG, Brazil  }
\date{}

\newcommand{\be}{\begin{equation}}
\newcommand{\ee}{\end{equation}}
\newcommand{\beq}{\begin{eqnarray}}
\newcommand{\eeq}{\end{eqnarray}}
\newcommand{{\xis}}{{\bf x}}
\newcommand{\ips}{{\bf y}}

\begin{document}
\maketitle
\thispagestyle{empty}
\begin{abstract}
Considering anyonic oscillators in a two-dimensional lattice, we realize  the
quantum semi-group $sl_{(q,s)}(2)$ by means of a generalized Schwinger
construction. We find that the parameter $q$ of the algebra is connected to the
statistical parameter, whereas the $s$ parameter is related to a $s$-deformed
oscillator introduced at each point of the lattice.
\end{abstract}

\newpage
\setcounter{page}{1}
%1111111111111111111111111111111111111111111111111111111111111111111111111111
%
%                             INTRODUCTION
%
%1111111111111111111111111111111111111111111111111111111111111111111111111111
\section{Introduction}

Quasitriangular Hopf Algebras, also called Quantum Groups
\citet{Drinfeld,Fadeev}, have attracted a lot of attention from physicists and
mathematicians in the last years. They have  found applications in several
areas of physics, such as: the inverse scattering method, vertex models,
anisotropic spin chains Hamiltonians, knot theory, conformal field theory,
heuristic phenomenology of deformed molecules and nuclei, non-commutative
approach to quantum gravity and anyon physics \cite{Zachos} (references
therein) and  \citet{Majid,Nos2}.

In the last case, an interesting connection between the quantum envelopping
algebras $sl_{q}(2)$ and anyons  \citet{Leinaas,Lerda2} was found \cite{Lerda}.
It was shown to be possible to realize
 $sl_{q}(2)$ via  a generalized Schwinger construction \cite{Schwinger}, using
non-local, intrinsically two-dimensional objects. These anyonic oscillators
are defined on a square lattice $\Omega$ and  interpolate between bosonic and
fermionic oscillators. The analysis of all deformed classical Lie algebra
\citet{Caracciolo,Frau} was also done, and the deformation parameter $q$ is
related to the statistical parameter $\nu$ as $q = \exp (i \pi \nu)$ ($q =\exp
( 2 i \pi \nu)$ for ${\cal U}_q(C_n)$).

These anyonic oscillators are hard core objects, and should not be confused
with  $q$-oscillators, since these objects are local and can live in any
dimension. The connection of $q$-oscillators  with quantum algebras was
recently investigatof the thermal properties of systems with quantum group
symmetry
\citet{Delgado,Monteiro} and the analysis of the possible application to
physical phenomena.

The aim of this letter is to construct the $sl_{(q,s)}(2)$ \cite{Burdik,Jing}
algebra with these anyonic oscillators. In the next section we review the main
results concerning anyonic oscillators, in section 3 we follow ref.\cite{Frau}
to construct a non-local set of generators for $sl_{(q,s)}(2)$. Section 4 is
devoted to constructing  anyonic oscillators and to showing how they are
connected to the generators, introduced in section 3, by the Schwinger method.
Then we shall also see that in order to realize  $sl_{(q,s)}(2)$  with anyonic
oscillators we had to  introduce a sort of $s$-oscillator at each point of the
lattice $\Omega$. We make some final remarks in the conclusion.

%2222222222222222222222222222222222222222222222222222222222222222222222222222
%
%                LATTICE ANGLE FUNCTIONS AND ANYONIC OSCILLATORS
%
%2222222222222222222222222222222222222222222222222222222222222222222222222222

\section{Lattice Angle Functions and Anyonic Oscillators}

In this section we are going to review the construction of  anyonic
oscillators defined on a two-dimensional square lattice $\Omega $ of spacing
one as  has been done in ref.\cite{Lerda}.

Anyonic oscillators are intrinsically non-local two-dimensional objects \citet
{Frad,Elie} which interpolate between fermionic and bosonic oscillators, that
can be constructed on a square lattice $\Omega $ by means of a Jordan-Wigner
\cite{Jordan} construction which in our case transmutes fermionic
oscillators into  anyonic ones.

To each point ${\bf x}=(x_1,x_2)$ of the lattice $\Omega $ we
associate a cut $\gamma _x$, made of bonds of the dual lattice $
\widetilde{\Omega }$ from minus infini(horizontal) axis, $o^{*}=(\frac 12,\frac
12)$ being the origin of the dual
lattice $\widetilde{\Omega}$. We denote by ${\bf x}_\gamma $ the point ${\bf
x}$ and its
associated cut $\gamma _x$.

The an\-gle func\-tion be\-tween two dif\-ferent points  ${\xis}$ and ${\ips}$
belonging to the lattice  $\Omega $ is de\-noted by $%
\Theta _{\gamma _x}({\bf x},{\bf y})$ and is defined as the angle of the point
${\xis}$
measured from the point $y^{*}$ belonging to $\widetilde{\Omega }$  with
respect to a line parallel to the positive $x$-axis.

One can show that
\begin{equation}
\label{teta}{\Theta }_{\gamma _x}({\bf x},{\bf y})-{\Theta }_{\gamma
_y}({\ips},{\xis})=\left\{
\begin{array}{c}
\pi \; sgn(x_2,y_2)
\mbox{~~ for~ }x_2\neq y_2 \\
\pi \; sgn(x_1,y_1) \mbox{~~ for~ }x_2=y_2.
\end{array}
\right.
\end{equation}
In fact, to arrive at this result it is necessary to neglect a term that
depends on the distance between $\xis$ and $\ips$, vanishing when they are far
apart. This typical lattice feature can be eliminated by embedding $\Omega $
into a lattice $\Lambda $ whose lattice spacing $\epsilon $ is much smaller
than $1$, then all quantities defined on $\Omega $ can be seen as
restrictions to $\Omega $ of quantities defined on $\Lambda $. The above
result is obtained when we let $\epsilon $ go to zero, since in this
limit all the points $\xis$ and $\ips$ of $\Omega $ are far apart, from the
point
of view of the lattice $\Lambda .$

Eq.(\ref{teta}) can be used to endow $\Omega $ with an ordering, which will
be very useful when dealing with anyonic oscillators. One chooses the
positive sign in eq.(\ref{teta}) and then one defines
\begin{equation}
\label{xmaiory}{\xis}>{\ips}=\left\{
\begin{array}{c}
x_2>y_2 \\
x_2=y_2,x_1>y_1,
\end{array}
\right.
\end{equation}
and eq(\ref{teta}) becomes
\begin{equation}
\labe_y}({\ips},{\xis})=\pi \mbox{~~~~~for ~} {\bf x} > {\bf y}.
\end{equation}

Even if unambiguous, this theta angle function introduced is not unique, as
it depends on the particular choice of the cuts $\gamma $. Another fundamental
choice can be obtained if we consider the cut $\delta$ made with bonds on
$\widetilde\Omega$ from plus infinity to $^{*}x=y-o^{*}$ parallel to the
$x$-axis \cite{Lerda}. With this cut $\delta$ one can define another lattice
angle  ${\widetilde{\Theta}}_{\delta_{x}}(\xis,\ips)$ which is the angle of
$\xis$ as seen from $^{*}y=y-o^{*}$. With the ordering given by
eq.(\ref{xmaiory}) it can be shown that
\begin{equation}
\label{teta3}\overline{{ \Theta }}_{\delta _x}({\bf x},{\bf
y})-\overline{{\Theta
}}_{\delta _y}({\ips},{\xis})=-\pi { \mbox{ ~~~~~ for ~} {\bf x} > {\bf y}}.
\end{equation}

One can also get from their definitions, a relation between these two angle
functions
\begin{equation}
\overline{ \Theta }_{\delta _x}({\bf x},{\bf y})-\Theta _{\gamma_x}({\bf
x},{\bf y})=\left\{
\begin{array}{r}
- \pi \mbox{~~~~ for ~} {\bf x} > {\bf y}  \\
\pi \mbox{~~~~~ for ~} {\bf x} < {\bf y},
\end{array}
\right.
\end{equation}
and for any $\xis$ and $\ips$ (even if $\xis=\ips$) one has
\begin{equation}
\overline{{ \Theta }}_{\delta _y}({\ips},{\xis})- {\Theta} _{\gamma_x}({\bf
x},{\bf y})=0.
\end{equation}

 One can use the theta function introduced above  to define anyonic
oscillators,
which are related by a parity transformation. One defines them as follows:
\begin{equation}
\label{anions}a_i({\xis}_\alpha )=K_i(\xis_\alpha )c_i(\xis),
\end{equation}
with $\alpha _x=\gamma _x$ or $\delta _x$, $i=1,...,N$; the disorder operator
given by
\be
K_i(x_\alpha )=\exp ( i\rho \sum_{\stackrel{\ips \in \Omega}{ \ips \neq \xis}}
\Theta _{\alpha _x}({\bf x},{\bf y})c_i^{\dagger}({\ips})c_i({\ips}) ) ,
\ee
and $c_i(x)$ are fermionic oscillators obeying
\be
\begin{array}{c}
\left\{ c_i({\xis}),c_i({\ips})\right\} =0 \\
\left\{ c_i({\xis})^{\dagger },c_i({\ips})\right\} =\delta _{ij} \; \delta
({\bf x},{\bf y}),
\end{array}
\ee
and their hermitean conjugate counterparts (which are going to be omitted
always in this letter). In the above formula $\delta ({\bf x},{\bf y})$ is the
delta function
on $\Omega $, i.e.
\be
\delta ({\bf x},{\bf y})=\left\{
\begin{array}{c}
0 \mbox{~~~ if~} {\bf x} \neq {\bf y} \\
1 \mbox{~~~ if~} {\bf x} = {\bf y}.
\end{array}
\right.
\ee

The anyonic oscillators of type $\gamma$ obey the following generalized
commutation relations for ${\bf x} > {\bf y}$
\beq
a_i({\xis}_{\gamma}) a_i({\ips}_{\gamma}) + q^{-1} a_i({\ips}_{\gamma})
a_i({\xis}_{\gamma}) = 0 \nonumber \\
a_i({\xis}_{\gamma}) {a_i}^{\dagger}({\ips}_{\gamma}) + q
{a_i}^{\dagger}({\ips}_{\gamma}) a_i({\xis}_{\gamma}) = 0,   \label{axy}
\eeq
where $q=\exp( i \pi \rho)$. For ${\bf x} = {\bf y}$ one has
\beq
&(a_i({\xis}_{\gamma}) )^2 = 0, & \nonumber \\
&a_i({\xis}_{\gamma}) {a_i}^{\dagger}({\xis}_{\gamma}) +
{a_i}^{\dagger}({\ips}_{\gamma}) a_i({\xis}_{\gamma}) = 1.& \label{aiai}
\eeq

 Thus, as one can see from the above discussion, anyonic oscillators are hard
core objects which obey $q$-commutation relations at different points of the
lattice but standard anticommutation relations at the same point.

%Two different anyonic oscillators atincommute, i.e.,
%\beq
%&\left\{ a_i({\xis}_{\gamma}) , {a_j}({\ips}_{\gamma}) \right\} = 0,&
%%%\nonumber \\
%&\left\{ {a_i}^{\dagger}({\ips}_{\gamma}) , a_j({\xis}_{\gamma}) \right\} =
%%%0.&  \label{aij}
%\eeq

The commutation relation among the anyonic oscillators of type $\delta$ can be
obtained from eqs.({\ref{axy}-{\ref{aiai}) by replacing $q$ by $q^{-1}$ and of
course the cuts $\gamma$ by $\de$\delta$-oscillators can be obtained from type
$\gamma$-oscillators by a parity
transformation which, as is well known, changes the braiding parameter $q$ to
$q^{-1}$ \cite{Lerda2}.

Commutaion relatins among different types of oscillators can also be computed,
and they read
\beq
&\left\{ a_i({\xis}_{\gamma}), {a_i}({\ips}_{\delta}) \right\} = 0 & \nonumber
\\
&\left\{ {a_i}^{\dagger}({\ips}_{\delta}) , a_i({\xis}_{\gamma}) \right\} = 0,&
\label{14}
\eeq
and
\be
\left\{ a_i({\xis}_{\delta}) , {a_i}^{\dagger}({\ips}_{\gamma}) \right\} = q^{
\left( \sum_{{\bf y}<{\bf x}} - \sum_{{\bf y>}{\bf x}} \right)
 {c_i}^{\dagger}({\bf y}) c_i({\bf y)}  }.  \label{anyslq}
\ee
Finally, we should mention that different anyonic oscillators (those made up of
different types of fermions) anticommute.
With the above defined anyonic oscillators one can realize all the classical
deformed algebras (with an introduction of a background term in the disorder
operator $K_i({\xis}_{\alpha})$ for the $B$ and $D$ series)
\cite{Caracciolo,Frau}.

%3333333333333333333333333333333333333333333333333333333333333333333333333333
%
%THE $sl_{(Q,S)}(2)$ QUANTUM SEMI-GROUP AND ITS NON-LOCAL REALIZATION
%
%3333333333333333333333333333333333333333333333333333333333333333333333333333
\section{The $sl_{(q,s)}(2)$ Quantum Semi-Group and its Non-Local Realization}

The commutation relations among the generators of the two-parametric quantum
algebra $sl_{(q,s)}(2)$ \cite{Burdik}
\beq \label{slqs2}
&[j_0,j_\pm] = \pm j_\pm,&  \nonumber \\
&{[}j_+,j_-{]}_s \equiv s^{-1} j_+ j_- - s j_- j_+ = s^{-2j_0} [2j_0],&
\eeq
where $[x] = \frac{q^{x} -q^{-x}}{q-q^{-1}}$, can be derived from the $R$
matrix \cite{Hlavaty}
\begin{equation}
R = \left(
\begin{array}{cccc}
q&0&0&0  \\
0&s&0&0 \\
0&q-q^{-1}&s^{-1}&0 \\
0&0&0&q
\end{array}
\right)
\end{equation}
wh\be
R_{12}R_{13}R_{23} = R_{23}R_{13}R_{12}.
\ee

The comultiplication structure of the algebra  \cite{Burdik}
\beq
&\Delta (qs)^{-j_0} = (qs)^{-j_{0}} \otimes (qs)^{-j_0}, &\nonumber \\
&\Delta (j_{\pm}) =  (qs)^{-j_0} \otimes j_{\pm} + j_{\pm} \otimes
{(qs^{-1})}^{j_0},&   \label{coproduto}
\eeq
together with the compatibility equations convert $sl_{(q,s)}(2)$ into a
bialgebra. It is not possible to find an antipode function for this algebra,
and thus $sl_{(q,s)}(2)$ is more properly  called a quantum semi-group. In the
limit $s \rightarrow 1$, $sl_{(q,s)}(2)$ goes to $sl_{(q)}(2)$.

An important fact about  $sl_{(q,s)}(2)$  is that  Pauli matrices are its
two-dimensional representation, thus the fundamental representation is the same
as for the $sl(2)$ algebra,
and all its  representations can be obtained from the fundamental one by the
use of the comultiplication rules given by
eq.(\ref{coproduto}).

Let us now go back to the lattice $\Omega$ introduced in the last section and
assign to each point $\bf x \in \Omega$ a fundamental representaion of
$sl_{(q,s)}(2)$ , its generators satisfying the local algebra
\beq
[j_0({\bf x }),j_\pm({\bf x} )] = \pm j_\pm(\bf x )  \nonumber \\
{[}j_+({\bf x }),j_-({\bf x }){]_s}= s^{-2j_0({\bf x} )} [2j_0({\bf x} )].
\label{20b}
\eeq
As the fundamental representation of  $sl_{(q,s)}(2)$  is the same as
$sl(2)$, the $q$-deformed structure of this equation is only formal, thus we
just write them in this way for future use.

With the local generators $j_0({\xis}),j_\pm({\xis})$ one can define
\beq
&J_0({\bf x}) = \prod_{{\bf y} < {\bf x}}^{\otimes} {\bf 1}_{\bf y} \otimes
j_0({\bf x}) \otimes  \prod_{{\bf z} > {\bf x}}^{\otimes}
{\bf 1}_{\bf z}&  \nonumber \\
&{J}_\pm({\bf x}) = \prod_{{\bf y} < {\bf x}}^{\otimes} {(qs)}^{-j_0{(\bf y)}}
\otimes
 j_\pm({\bf x}(\bf z)},&
\eeq
(hereafter we shall drop the symbol of the direct product) and the generators
\beq
J_{\pm} = {\sum_{{\xis} \in \Omega}} J_{\pm}({{\xis}}) \nonumber \\
J_{0} = {\sum_{{\xis} \in \Omega}} J_{0}({{\xis}}),  \label{global}
\eeq
obey the algebra of  $sl_{(q,s)}(2)$ , eq.(\ref{slqs2}), as they are obtained
by the iterated coproduct of the envelopping algebra.

The generators  $ J_{0}({\xis}), J_{\pm}({\xis})$  defined above obey the
commutation relations
\beq
&[J_0({\bf x} ),J_\pm({\bf y} )] = \pm \delta{({\xis},\ips)} \; J_\pm({\bf x} )
 & \nonumber \\
&{[}J_+({\bf x} ),J_-({\bf y} ){]} = 0   {\mbox{~~~~}}\xis \neq \ips &
\nonumber \\
&\left[ J_+({\xis}), J_-({\ips}) \right]_s = \displaystyle{\prod_{{\bf y} <
{\bf x}}}
 {(qs)}^{-2j_0{({\bf y})}}  [ j_+({\xis}), j_-({\ips}) ]_s
\displaystyle{\prod_{{\bf z} > {\bf x}}} {(qs^{-1})}^{2 j_0 ({\bf z})},&
\label{21}
\eeq
and the densities $J_\pm({\xis})$ obey the braiding relations
\beq
 J_+({\xis}) J_+({\ips}) = q^2  J_+({\ips}) J_+({\xis}) \nonumber  \\
 J_-({\xis}) J_-({\ips}) = q^{-2}  J_-({\ips}) J_-({\xis}),  \label{23}
\eeq
which could be used to prove directly that $J_0,J_\pm$ obey the $sl_{(q,s)}(2)$
algebra eq.(\ref{slqs2}).

Let us now use the angles $\Theta_{\gamma_x}({\xis},{\ips})$ and
$\bar{\Theta}_{\delta_x}({\xis},{\ips})$ introduced in the last section to
construct new non-local densities $J_0({\xis})$, $J_\pm({\xis})$
\beq
& J_+({\xis}) = \displaystyle{\prod_{{\bf y} < {\bf x}}} {q}^{-\frac{2}{\pi}
\theta_{\gamma_{\xis}}({\xis},{\ips}) j_0({\ips})} \; s^{-2j_0({\ips})} \;
j_+({\xis})
\displaystyle{\prod_{{\bf z} > {\bf x}}} {q}^{-\frac{2}{\pi}
\theta_{\gamma_{\xis}}({\xis},{\bf z}) j_0({\bf z})} \; s^{-2j_0({\bf z})}&
\nonumber \\
&J_-({\xis}) = \displaystyle{\prod_{{\bf y} < {\bf x}}} q^{\frac{2}{\pi}
\bar{\th\; j_-({\xis})
\displaystyle{\prod_{{\bf z} > {\bf x}}} {q}^{\frac{2}{\pi}
\bar{\theta}_{\delta_{\xis}}({\xis},{\bf z}) j_0({\bf z})} \;  s^{-2j_0({\bf
z})}&  \nonumber  \\
&J_0({\xis}) = \displaystyle{\prod_{{\bf y} < {\bf x}}} 1_{\ips} \; j_0({\xis})
\;
\displaystyle{\prod_{{\bf z} > {\bf x}}} 1_{\bf z}. \label{jteta}&
\eeq

Using the relations obeyed by the theta-angle functions and the local algebra
eq.(\ref{20b}), we can prove that these densities obey the commutation
relations eq.(\ref{21}) as well as the braiding relations eq.(\ref{23}) and
then realize the algebra $sl_{(q,s)}(2)$, eq.(\ref{slqs2}).

%4444444444444444444444444444444444444444444444444444444444444444444444444444
%
%                      ANYONIC REALIZATION OF $sl_{(Q,S)}(2)$
%
%4444444444444444444444444444444444444444444444444444444444444444444444444444

\section{Anyonic Realization of $sl_{(q,s)}(2)$}

In this section we are going to show that the anyonic oscillators defined in
section $1$, with a suitable choice of the disorder operator
$K_i({\xis}_{\alpha})$, realize, via a Schwinger like construction the algebra
of densities eq.(\ref{21}) and the braiding relations eq.(\ref{21}), and
consequently also the $sl_{(q,s)}(2)$ algebra eq.(\ref{slqs2}).

We begin by recalling that all classical Lie algebras can be constructed {\it
\`a la} Schwinger on a manifold $\Omega$ in terms of fermionic oscillators. In
particular, for the  $sl(2)$ algebra one can define at each point ${\xis}$ of
$\Omega$
\beq
&j_+({\xis}) = {c_1}^{\dagger}({\xis}) c_1({\xis}) & \nonumber \\
&j_0({\xis}) = \frac{1}{2} \left( {c_1}^{\dagger}({\xis}) c_1({\xis}) -
{c_2}^{\dagger}({\xis}) c_2({\xis}) \right) & \nonumber \\
&j_-({\xis}) = {c_2}^{\dagger}({\xis}) c_2({\xis}),&
\eeq
where $c_i({\xis}) $ are fermionic oscillators.
These operator\beq
&[j_0({\bf x }),j_\pm({\bf y} )] = \pm \, \delta{({\xis},\ips)} \, j_\pm({\bf
x} )  & \nonumber \\
&{[}j_+({\bf x} ),j_-({\bf y} ){]} = 2 \, j_0 ({\xis}) \,
\delta{({\xis},\ips)}.  & \label{28}\eeq

Once more, global generators $J_\pm$, $J_0$ can be defined from the densities
$J_\pm({\xis})$, $J_0({\xis})$,
\beq
J_{\pm} = {\sum_{{{\xis}} \in \Omega}^{}} J_{\pm}({{\xis}}) \nonumber \\
J_{0} = {\sum_{{{\xis}} \in \Omega}^{}} J_{0}({{\xis}}),
\eeq
where
\beq
&J_0({\bf x}) = \displaystyle{\prod_{{\bf y} < {\bf x}}} {\bf 1}_{\bf y} \;
j_0({\bf x})  \; \displaystyle{\prod_{{\bf z} > {\bf x}}}
{\bf 1}_{\bf z}&  \nonumber \\
&J_\pm({\bf x}) = \displaystyle{\prod_{{\bf y} < {\bf x}}} {\bf 1}_{\bf y} \;
j_0({\bf x}) \;   \displaystyle{\prod_{{\bf z} > {\bf x}}}
{\bf 1}_{\bf z},&
\eeq
and it is very easy to see that  $J_\pm$, $J_0$ closes under $sl(2)$. The
spin-$0$ and spin-$1/2$ representations of the local algebra can be combined to
give all the unitary representations of $sl(2)$.

As  was stated in the last section, the $sl_q(2)$ algebra can also be generated
by this local $sl(2)$ algebra, just by changing the comultiplication rules for
the representations. From the point of view of the Schwinger construction, this
is equivalent to changing the oscillators of eq.(\ref{28}) into the anyonic
oscillators introduced in section 2. In fact, with the choices of densities
\beq
&J_+({\xis}) = {a_1}^{\dagger}({\xis_\gamma}) a_2({\xis_\gamma})& \nonumber \\
&J_0({\xis}) = \frac{1}{2} \left( {a_1}^{\dagger}({\xis_\gamma})
a_1({\xis_\gamma}) - {a_2}^{\dagger}({\xis_\gamma}) a_2({\xis_\gamma}) \right)
& \nonumber \\
&j_-({\xis}) = {a_2}^{\dagger}({\xis_\delta}) a_1({\xis_\delta}),&
\eeq
and with the help of eq.(\ref{anyslq}) it is possible to see that
$J_\pm({\xis})$, $J_0({\xis})$, obey eq.(\ref{21}-\ref{23}) for $s=1$
\cite{Lerda}densities will obey the $sl_q(2)$ algebra. We notice here that the
choice of
the cut $\gamma$ in $J_0$ is immaterial, since the product
${a_i}^{\dagger}({\xis_\alpha}) a_i({\xis_\alpha})$ can be written in terms of
fermionic oscillators without any dependence on the disorder operator
$K_i({\xis}_{\alpha})$.

The Schwinger construction of  $sl_{(q,s)}(2)$ algebra has, however, a subtlety
due  to the presence of the $s$-commutator in the local algebra eq.(\ref{20b}).
Let us now define the local generators
\beq
&j_+({\xis}) = {c_1}^{\dagger}({\xis}) s^{-\frac{1}{2}
({c_1}^{\dagger}({\xis}) c_1({\xis}) - {c_2}^{\dagger}({\xis}) c_2({\xis}))}
c_1({\xis})& \nonumber \\
&j_0({\xis}) = \frac{1}{2} \left( {c_1}^{\dagger}({\xis}) c_1({\xis}) -
{c_2}^{\dagger}({\xis}) c_2({\xis}) \right) &\nonumber \\
&j_-({\xis}) = {c_2}^{\dagger}({\xis}) s^{\frac{1}{2} ({c_1}^{\dagger}({\xis})
c_1({\xis}) - {c_2}^{\dagger}({\xis}) c_2({\xis})) } c_1({\xis}).&
\eeq
It is easy to see that these local generators close under the algebra
eq.(\ref{20b}).

The anyonic oscillators can be defined as
\be
A_i({\xis}_\alpha) = K_i ({\xis}_\alpha) b_i({\xis_\alpha}), \label{anions2}
\ee
with
\beq
K_i({\xis}_\gamma)=  \exp^{ \; \;  \displaystyle{\sum_{{\ips} \neq {\xis}}}
\left( i \rho  \Theta _{\gamma _x}({\bf x},{\bf
y})c_i^{\dagger}({\ips})c_i({\ips}) + \frac{i}{2} \nu \pi
c_i^{\dagger}({\ips})c_i({\ips}) \right) }  \nonumber \\
K_i({\xis}_\delta)= \exp^{ \; \; \displaystyle{\sum_{ {\ips} \neq {\xis}}}
\left( i \rho  \bar{\Theta} _{\delta _x}({\bf x},{\bf
y})c_i^{\dagger}({\ips})c_i({\ips})  - \frac{i}{2} \nu \pi
c_i^{\dagger}({\ips})c_i({\ips}) \right) },  \label{disorder2}
\eeq
and
\beq
b_i(\xis_{\gamma}) = \exp^{(\frac{ i \nu \pi}{2} {c_i}^{\dagger}(\xis)
c_i(\xis))} c_i(\xis) \nonumber \\
b_i(\xis_{\delta}) = \exp^{(\fc_i(\xis))} c_i(\xis). \label{b1}
\eeq
These operators obey, at each point ${\xis} \in \Omega$, the algebra
\beq
&{b_i}^2(\xis_\alpha) = 0& \nonumber \\
&\{ b_i(\xis_\alpha), {b_i}^{\dagger}(\xis_\alpha) \} = 1&  \nonumber \\
&\{ b_i({\xis}_\alpha), {b_j}^{\dagger}({\xis}_\alpha) \} = 0  \mbox{~~ $i\neq
j$},&  \label{b2}
\eeq
where $\alpha$ can be $\gamma$ or $\delta$, and also
\be
 b_i({\xis}_\gamma) {b_i}^{\dagger}({\xis}_\delta) +  s \;
{b_j}^{\dagger}({\xis}_\delta)  b_i({\xis}_\gamma)  = s^{ N_i ({\xis})},
\label{b3}
\ee
where $N_i =  {c_i}^{\dagger} c_i$. So the $b$ operators are hard-core objects
that obey a $s$-deformed Heisenberg algebra at each point of the lattice
$\Omega$.

 With this choice, the densities $J_\pm({\xis})$, $J_0({\xis})$  defined by
\beq
&J_+({\xis}) = {A_1}^{\dagger}({\xis_\gamma}) A_2({\xis_\gamma})& \nonumber \\
&J_0({\xis}) = \frac{1}{2} \left( {A_1}^{\dagger}({\xis_\gamma})
A_1({\xis_\gamma}) - {A_2}^{\dagger}({\xis_\gamma}) A_2({\xis_\gamma}) \right)
& \nonumber \\
&J_-({\xis}) = {A_2}^{\dagger}({\xis_\delta}) A_1({\xis_\delta}),&
\eeq
obey the commutation relations eq.({\ref{21}), the braiding relations
(\ref{23}) and consequently, $J_0,J_\pm$ defined by
\beq
J_{\pm} = {\sum_{{{\xis}} \in \Omega}^{}} J_{\pm}({{\xis}}) \nonumber \\
J_{0} = {\sum_{{{\xis}} \in \Omega}^{}} J_{0}({{\xis}}),
\eeq
satisfy the algebra of $sl_{(q,s)}(2)$.

 From their definition, it is easy to see that the disorder operators
$K_i(\xis_\alpha)$ commute among themselves
\be
K_i(\xis_\alpha) \; K_j(\ips_\beta) = K_j(\ips_\beta) \; K_i(\xis_\alpha),
\mbox{~~~~ for all $\xis$,$\ips$}  \label{kcomk}
\ee
for any value for $i,j$, where the cuts $\alpha$ and $\beta$ can be either
$\gamma$ or $\delta$,

 Eq.(\ref{kcomk}), together with eq.(\ref{b2},\ref{b3}) give the following
relations for the operators $\beq
A_i(\xis_\gamma)  A_i(\ips_\gamma) = - q^{-1} A_i(\ips_\gamma) A_i(\xis_\gamma)
\nonumber \\
A_i(\xis_\gamma) {A_i}^{\dagger}(\ips_\gamma) = - q A_i(\ips_\gamma)
A_i(\xis_\gamma),
\eeq
for all $\xis > \ips$. At the same point we have
\be
\left\{ A_i(\xis_\gamma),{A_i}^{\dagger}(\xis_\gamma) \right\} = 1.
\ee

The relations for the cut $\delta$ can be obtained by changing $q$ into
$q^{-1}$ in the relations above. We can also find relations among oscillators
defined with different cuts, and they read
\be
A_i(\xis_\gamma)  A_i(\ips_\delta) = -s^{-1} A_i(\ips_\delta) A_i(\xis_\gamma),
\ee
for all $\xis$,$\ips$ and
\be
A_i(\xis_\delta)  {A_i}^{\dagger}(\ips_\gamma) = - s^{-1}
{A_i}^{\dagger}(\ips_\gamma) A_i(\xis_\delta),
\ee
for $\xis \neq \ips$. If $\xis = \ips$ we have
\beq
A_i(\xis_\gamma) {A_i}^{\dagger}(\xis_\delta) &=&  K_i ({\xis}_\gamma)
b_i({\xis_\gamma}), {K_i}^{\dagger}({\xis}_\delta)
{b_i}^{\dagger}({\xis_\delta})  \nonumber \\
&=& {K_i}^{\dagger}({\xis}_\delta) b_i({\xis_\gamma})
{b_i}^{\dagger}({\xis_\delta})  K_i ({\xis}_\gamma) \nonumber \\
&=& {K_i}^{\dagger}({\xis}_\delta) \left( s^{N_i(\xis)} - s
{b_i}^{\dagger}({\xis_\delta})  b_i({\xis_\gamma})  \right)  K_i
({\xis}_\gamma) \nonumber \\
&=&-s  {A_i}^{\dagger}(\xis_\delta) A_i(\xis_\gamma) +  s^{N_i(\xis)}
{K_i}^{\dagger}({\xis}_\delta) K_i ({\xis}_\gamma),  \nonumber \\
\eeq
implying the relation
\be
A_i(\xis_\gamma) {A_i}^{\dagger}(\xis_\delta) + s
{A_i}^{\dagger}(\xis_\delta) A_i(\xis_\gamma) =  q^{
\left( {\displaystyle \sum_{{\bf y}<{\bf x}}} - {\displaystyle{\sum_{\ips
>\xis}}} \right)
 {c_i}^{\dagger}({\bf y}) c_i({\bf y)}  }  s^{\;\displaystyle{\sum_{\ips}}
N(\ips)}.
\ee

The construction we have performed in this section for $sl_{(q,s)}(2)$ algebra
uses anyonic oscillators made with $s$-deformed oscillators. This is a special
characteristic of the twoalgebras were constructed with anyonic oscillators
built up of fermions
\cite{Lerda,Caracciolo,Frau}.
%5555555555555555555555555555555555555555555555555555555555555555555555555555
%
%                               CONCLUSIONS
%
%5555555555555555555555555555555555555555555555555555555555555555555555555555

\section{Conclusions}

In this letter we have realized the quantum semi-group $sl_{(q,s)}(2)$ on a
two-dimensional lattice $\Omega$. We have first showed that the generators can
be written as a non-local expression  made of the lattice angle function on
$\Omega$ and we have discussed its connection with a Schwinger like
construction using anyonic oscillators.

Differently from the cases previously considered \cite{Lerda,Caracciolo,Frau},
we had to consider a different kind of anyonic oscillators. These anyonic
oscillators are made with $s$-deformed fermionic oscillators defined on
$\Omega$, instead of pure fermionic oscillators. The parameter $q$, as in the
previous cases, is connected to the statistical parameter $\nu$ by
$q=\exp(i\pi\nu)$.

We find that it would be interesting to generalize the analysis we have
performed for the case of multiparametric deformed algebras in order to see
what would be the role of the various parameters, as in the case of
two-parameter algebra  one being associated to the statistical parameter and
the other being related to the parameter of the deformed Heisenberg algebra.

The  connection of $q$-oscillators with quantum algebras permitted the
investigation of the possible applications of quantum groups  to physical
problems through the analysis of the thermal properties of deformed systems
\citet{Delgado,Jing}. We consider that the analysis of the connection between
quantum algebras and anyons could, as well, provide another area of research on
the possible applanyonic interpretation of planar physics.

\vspace{1cm} {\bf Acknowledgement:} One of the authors (JLMV) wish to thank
CNPq for financial support.
\newpage

%BBBBBBBBBBBBBBBBBBBBBBBBBBBBBBBBBBBBBBBBBBBBBBBBBBBBBBBBBBBBBBBBBBBBBBBBBBB
%
%                            BIBLIOGRAFIA
%
%BBBBBBBBBBBBBBBBBBBBBBBBBBBBBBBBBBBBBBBBBBBBBBBBBBBBBBBBBBBBBBBBBBBBBBBBBBB

\end{document}